\documentclass{article}
\usepackage[utf8]{inputenc}
\usepackage{amsmath}
\usepackage{graphicx}
\usepackage[ruled,vlined]{algorithm2e}
\usepackage{bbm}
\usepackage{authblk}
\usepackage{amsfonts}

\title{Faster Regret Matching}
\author{Dawen Wu}
\affil{Peking University}
\date{December 2019}

\begin{document}

\maketitle
\begin{abstract}
    The regret matching algorithm proposed by Sergiu Hart is one of the most powerful iterative methods in finding correlated equilibrium. However, it is possibly not efficient enough, especially in large scale problems. We first rewrite the algorithm in a computationally practical way based on the idea of the regret matrix. Moreover, the rewriting makes the original algorithm more easy to understand. Then by some modification to the original algorithm, we introduce a novel variant, namely faster regret matching. The experiment result shows that the novel algorithm has a speed advantage comparing to the original one.
\end{abstract}
\section{Introduction}
\subsection{Game}

Game theory is a well-studied discipline that analysis situations of competition and cooperation between several involved players. It has a tremendous application in many areas, such as economic, warfare strategic, cloud computing. Furthermore, there are some studies about using the game theory to interpret machine learning model in recent years \cite{ref1,ref2}.

One of the milestones in game theory \cite{ref3,ref4,ref5} is, John von Neumann proved the famous minimax theorem for zero-sum games and showed that there is a stable equilibrium point for a two-player zero-sum game \cite{ref6}. Later, another pioneer, John Nash proved that, at every n-player general sum game with a finite number of actions for each player, there must exist at least one Nash equilibrium.

We now give a formal math definition of a game. Let $\Gamma=\left(N,\left(S^{i}\right)_{i \in N},\left(u^{i}\right)_{i \in N}\right)$ be a finite action N-person game. $N$ is players set. $S^i$ is the set of actions of player $i$. In this paper, we consider pure strategy $s\in S$ particularly, which means each player chooses one action only. $u^i: S \rightarrow \mathbb{R}$ is the payoff function of player $i$, and, in the two-person finite action case, the payoff function of one player is a matrix with size $|S^1|*|S^2|$. The entry in this matrix, say $m^i_{j,k}$, means the payoff value of player $i$ given that player one chooses action $j$, and player two chooses action $k$. We will give more detail about the notations in the section 1.3. 

A Nash equilibrium \cite{ref7,ref8} of a game is a strategy profile when no player can get more benefit by unilaterally changing his strategy.  $s\in S$ is a Nash equilibrium, such that for each $i$, it satisfies 
\begin{equation}
u^i(s) \geq u^i(s^i=k;s^{-i}), \forall k \in S^i.
\end{equation}

\subsection{Correlated equilibrium}

Robert Aumann originally presents a solution concept called correlated equilibrium \cite{ref9,ref10,ref11}. It is more general than the well known Nash equilibrium. The correlated equilibrium is a public recommend strategy giving to each player. If every player gains no benefit of deviation, we say this recommend strategy a correlated equilibrium. The formal math definition is given below. If for every $i\in N$, every $j,k\in S^i$, a probability distribution $\psi$ satisfy:

\begin{equation}
\label{eq:1}
    \sum_{s \in S: s^{i}=j} \psi(s)\left[u^{i}\left(k, s^{-i}\right)-u^{i}(s)\right] \leq \alpha,
\end{equation}

 then the $\psi$ is a correlated $\alpha$-equilibrium.
Moreover, if $\alpha = 0$, $\psi$ is a correlated equilibrium. Instead of finding the Nash equilibrium of a game, in this paper, we want to find the correlated equilibrium.

\subsection{Notation}
In general, we use superscript to denote player index and subscript to denote time. $S^i$ means the action set of player $i$, and it contains $|S^i|$ numbers of elements. $S:=\Pi_{i \in N} S^{i}$ is the set of N-tuple of pure strategy. $s_t$ represents a action profile at time t, which is a $n*1$ vector, and the $i$-th element describes the $i$-th player action. $s^{-i}$ denote a $(n-1)*1$ size pure strategy which is exactly the same as $s$ except the $i$-th player has no action, i.e., $s^{-i}=\left(s^{i^{\prime}}\right)_{i^{\prime} \neq i}$. We denote $s^i=k;s^{-i}$ a pure strategy, which is the same as $s$ except the $i$-th player choose action $k$. This notation is widely used in the following chapter. $u^{i}: \prod_{i \in N} S^{i} \rightarrow \mathbb{R}$ is the player $i$ payoff function.

\subsection{Contributions}

Our contribution of this paper is threefold.
\begin{itemize}
    \item We bring up the concept of regret matrix, making the original algorithm more easy to implement and understand.
    \item We propose a variant of the regret matching by utilizing the negative part of the regret values. This novel algorithm is possibly faster than the original one.
    \item We develop an experiment to analyze both the original algorithm and the variant
\end{itemize}

\section{Rewriting the regret matching in a computationally friendly way}
The whole algorithm base on the adaptive procedure named regret matching proposed by Sergiu \cite{ref12,ref13, ref14,ref15,ref16,ref17,ref18}. Indeed, we rewrite the method of Sergiu in a computationally friendly way. The whole algorithm is essentially the same as regret matching, and it does hold the great convergence guarantee.

The first point to make is the separated player updating rule.  Each player has its own payoff function $u^i$, and the algorithm does not require any knowledge about other player's payoff functions. It updates its own regret-matrix and deriving probability distribution from it. That is, each player has completely nothing to deal with other players. The algorithm explained below is built on some fixed player $i$ and some fixed time $t$, and it can extend to all the other players $i$ and time $t$.

We can now give a brief introduction to the algorithm. At the time $t$, player $i$ in the game wants to find the probability distribution$p_t^i$ of its action set$S^i$, which will control the player action choice at the next time $t+1$. The probability distribution is derived from the regret-matrix$R^i$. So at time $t$, after receiving the payoff value, each player updates its regret-matrix base on the payoff value, and the regret-matrix subsequently infer a probability distribution controlling the next action.

Each player holds a so-called regret-matrix. The regret-matrix $R^i$ is a $|S^i|*|S^i|$ size matrix, where $|S^i|$ is the action set size of player $i$. The entry $r_{j,k}$ of regret-matrix $R^i$ represents the regret value up to the present time $t$, i.e., if the player $i$ changes the action $j$, which was played in the past, to the action $k$, how much more can it obtain. The subscript  $j$ means the action player $i$ chose at time $t$ and the other subscript $k$ means the player $i$ switch action $j$ to action $k$ in the whole playing history up to time t. At the time $t$, we fix a row $r_j$ corresponded to the action $j$ we have chosen in the last period, and the matrix updating operation and the probability distribution deriving operation are all happening in this $r_j$ row. Note that all the diagonal elements in the regret-matrix are zero because nothing will change if you change action $j$ to action $j$ itself.

\begin{figure}
    \centering
    \includegraphics[scale=0.5]{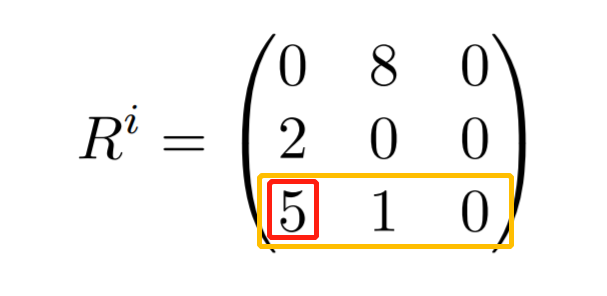}
    \caption{An Regret-Matrix Example}
    \label{An regret-matrix}
\end{figure}

Figure \ref{An regret-matrix} shows a regret-matrix example of player $i$ with three actions. The yellow box means that at time t player $i$ choose the third action. So at time t, player $i$ exclusively care about the third row of the matrix, and both the updating and deriving procedure happen on the third row $r_3$. The red box represents that player $i$ want to replace action three by action one in the whole history playing record. The number 5 in the red box means the player $i$ can get five more values if the replacing happens.

The core of the algorithm is divided into two parts, the regret-matrix updating, and the probability distribution deriving. Assuming the action profile is $s \in S$, the payoff of player i at time t is $u_t^i(s)$. The player $i$ uses this information solely to update its regret-matrix, more specifically the row $r_j$. The updating rule is 
\begin{equation}
\label{eq:2}
    r^i_{j,k} = r^i_{j,k} + u^i(s^i=k;s) - u^i(s),
\end{equation}
for all $k\in S^i$.

After updating the regret-matrix $R^i$, the next step is how to derive a probability distribution from it. The deriving formula is  deriving formula is 
\begin{equation}
\label{eq:3}
    p_{t+1}^i(k) = \frac{1}{\mu}*\frac{1}{t}*\max\{r^i_j[k],0\},
\end{equation}
for all $k\in A^i$ and $k \neq j$. For $ p_{t+1}^i(j)$, the formula is
\begin{equation}
\label{eq:4}
    p_{t+1}^{i}(j)=1-\sum_{k \in S^{i}: k \neq j} p_{t+1}^{i}(k)
\end{equation}

\begin{algorithm}[H]
\SetAlgoLined
\SetKwInOut{Input}{input}
\SetKwInOut{Output}{output}

\Input{Game, Initial probability distribution $p_0$}
\Output{Empirical playing record}
\For{t = 0,...,T-1}{
    Draw pure strategy $s_t$ according to the probability distribution $p_t$.\\
    \For{each player $i$}{
        compute $u^i(s_t)$ for updating the regret-matrix.\\
        update $r_j$ according to equation \ref{eq:2}, where $j=s_t$.\\
        derivate $p_{t+1}^i$ from $r_j$ according to equation \ref{eq:3}.
    }
}
\caption{Regret-Matching}
\end{algorithm}

As mentioned above, the whole algorithm control n regret-matrix, the total number of players, and each matrix has the size of $|S^i|*|S^i|$ corresponded to each player $i$. So in total, the algorithm involves $\sum_{i\in N} |S^i|^2$ numbers of elements. At each iteration, $\sum_{i\in N} |S^i|$ elements will change.

\section{Faster regret matching}
There is an interesting fact that most learning algorithms in finding the Nash equilibrium or correlated equilibrium use the non-negative regret value. When generating the probability distribution, the algorithm introduced in section two has a clip operation, which is to remove the negative part of regret leaving the positive part only. However, the negative parts do provide valuable information also, maybe not as much as the positive parts did. In order to accelerate the process of finding the correlated equilibrium, we utilize the negative regret value, leading to a potentially faster algorithm.

To be specific, we remove the max operation in equation \ref{eq:3} retaining the negative regret value. Nevertheless, it will lead to the vector p in the left not to add up to zero, so that it is not a probability distribution. To overcome this problem, we add the softmax operation after equation \ref{eq:3} and remove the equation \ref{eq:3} because, after softmax, the vector p is naturally a probability distribution. The softmax operation is 
\begin{equation}
\label{eq:5}
p_{t+1}^i(k)\xleftarrow[]{}\frac{e^{p_{t+1}^i(k)}}{\sum_{j=1}^{K} e^{p_{t+1}^i(j)}},
\end{equation}
for every $k \in S^i$.

This modification has effectively used the information of negative regret value, and guarantee the vector $p$ still a probability distribution.
Moreover, it makes the algorithm even more intuitive and straightforward.  In the following section, we experimentally show its efficiency.

Although this improvement provides speed increasing, however, unlike the original algorithm, it cannot guarantee the empirical distribution converge to correlated equilibrium. So there is a tradeoff between faster speed and convergence guarantee.

\section{Experiment}
We conduct the experiment using the i7-9700 processor, and the whole experiments are implemented in Python 3.7 and Numpy 1.16.5. The code is available on github.com.

\begin{figure}
    \centering
    \includegraphics[scale=0.5]{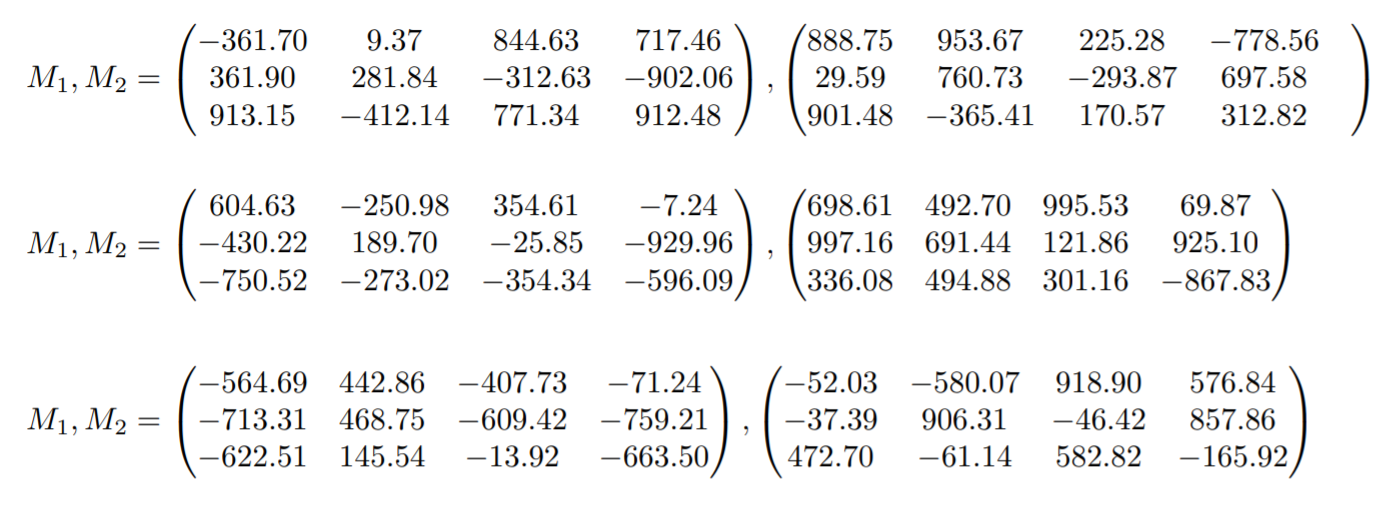}
    \caption{Three payoff matrices}
    \label{payoff matrix}
\end{figure}

For the sake of simplicity, we consider a two-person general sum game, and player one has three actions, player two has four actions. The payoff-matrix's elements of each player are sampled from the universal distribution ranging from $-1000$ to $1000$. We repeated the experiment three times. Figure  \ref{payoff matrix} shows the payoff-matrix used in the experiment.

We use the alpha value to measure the distance from the current empirical distribution to the correlated equilibrium point. The alpha value is defined in equation \ref{eq:1}. When the alpha value is equal to zero, it means the empirical distribution is a correlated equilibrium.

Let each player following the algorithm introduced in section two. The empirical distribution gradually converges to correlated equilibrium. In addition, if using the faster version of the algorithm proposed in section three, the alpha dorp down more quickly, which verifies our analysis. 

\begin{figure}[h]
    \centering
    \includegraphics[scale=0.4]{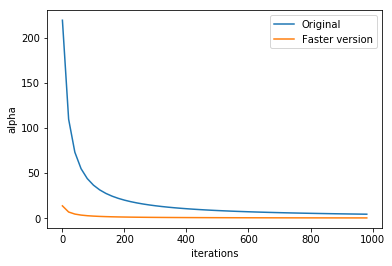}
    \includegraphics[scale=0.4]{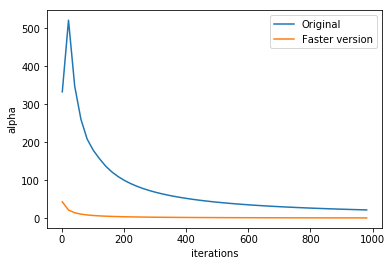}
    \includegraphics[scale=0.4]{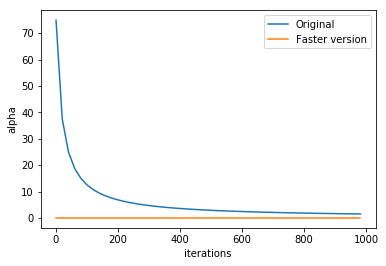}
    \caption{The learning procedure of two methods}
    \label{comparasion}
\end{figure}

Figure \ref{comparasion} illustrates the learning procedure of two methods. It base on the payoff function showed above. The total iteration number $T$ is 1000, and we measure the alpha value every 20 times.

\section{Conclusion}

We first review some concepts in game theory needed in this paper. Then we rewrite the regret matching algorithm for convenient implementing and easy understanding. The rewriting fundamentally bases on the regret-matrix. After the rewriting, we improve the original regret matching algorithm by utilizing the negative parts of the regret values, and the idea is intuitive and natural. The experiment result shows the efficiency of this improvement.

Moreover, there are some interesting points worth further discussing. The original regret matching algorithm owns the extraordinary convergence property. However, the convergence of the new method we proposed is not guaranteed. We hardly find a good way to ensure the convergence, which is the limit of our paper. Possibly good work is deeply analyzing the mechanism behind this new method, which may find out the reason for the un-convergence problem. Another possibly further work is using the negative parts of regret value in a more smart way, for more speed increasing and certainly converging.

\end{document}